\documentclass[article]{aastex63}
\usepackage{amsmath}
\usepackage{hyperref}
\usepackage{todonotes}
\usepackage{bm}
\usepackage{booktabs}
\usepackage{multirow}

\shorttitle{Non-simultaneous Regression Biases}
\shortauthors{Dixon}

\graphicspath{{./}{figures/}}

\begin{document}

\title{Biases from Non-Simultaneous Regression with Correlated Covariates: A Case Study from Supernova Cosmology}

\correspondingauthor{Samantha Dixon}
\email{sam.dixon@berkeley.edu}

\author[0000-0003-1861-0870]{Samantha Dixon}
\affiliation{Department of Physics, University of California Berkeley, 366 Physics North MC 7300, Berkeley, CA, 94720-7300}
\affiliation{Physics Division, Lawrence Berkeley National Laboratory, 1 Cyclotron Road, Berkeley, CA, 94720}

\newcommand{\sgn}{\text{sgn}}

\begin{abstract}
Several Type Ia supernova analyses make use of non-simultaneous regressions between observed supernova and host galaxy properties and supernova luminosity: first the supernova magnitudes are corrected for their light curve shape and color, and then they are separately corrected for their host galaxy masses. This two-step regression methodology does not introduce any biases when there are no correlations between the variables regressed in each correction step. However, correlations between these covariates will bias estimates of the size of the corrections, as well as estimates of the variance of the final residuals. In this work, we analyze the general case of non-simultaneous regression with correlated covariates to derive the functional forms of these biases. We also simulate this effect on data from the literature to provide corrections to remove these biases from the data sets studied. The biases examined here can be entirely avoided by using simultaneous regression techniques.
\end{abstract}

\keywords{Linear regression (1945); Multivariate analysis (1913); Type Ia supernovae (1728)}

\section{Introduction} \label{sec:intro}
Properties of Type Ia supernovae (SNe Ia) have been observed to be correlated with their absolute luminosities. Before accounting for these properties, the absolute brightnesses of typical SNe Ia vary by $\sim 0.4$ magnitudes. After accounting for correlations with the decay time of the light curve and the color of the object, their corrected absolute brightnesses are consistent to within $\sim 0.14$ mag \citep{phillips_absolute_1993, hamuy_morphology_1996, riess_precise_1996, perlmutter_measurements_1997, tripp_two-parameter_1998}. These calibrated brightness estimates make SNe Ia powerful cosmological distance indicators, and when combined with redshift measurements, allow us to map out the expansion history of the Universe. This technique was instrumental in the discovery of the accelerating expansion of the Universe \citep{perlmutter_measurements_1999, riess_observational_1998}, and continues to serve as a powerful probe of the nature of the dark energy driving this acceleration.

A common analysis method for standardizing supernova brightnesses uses the SALT2 spectral model \citep{guy_salt2_2007, betoule_improved_2014, mosher_cosmological_2014} to parametrize SN~Ia light curves. The model parameters represent an individual supernova's peak apparent brightness in the Bessell B-band ($m_B^*$), temporal width ($x_1$), and observed color ($c$). The distance modulus, $\mu$, to each object $i$ at redshift $z_i$ is then modeled as a linear combination of these parameters:
\begin{equation}
    \mu_i(z_i) = m_{B\;i}^*(z_i) - M + \alpha x_{1\;i} - \beta c_i
    \label{eqn:salt_mu}
\end{equation}
Typically, we would find the values of $M$, $\alpha$, and $\beta$ by minimizing the following quantity with respect to these parameters as well as the cosmological parameters of interest:
\begin{equation}
    \chi^2 = \displaystyle\sum_{i} \frac{\mu_i(z_i; m_{B, i}^*, x_{1,i}, c_i)-\mu_\text{cosmo}(z_i; \bm{\Theta})}{\sigma_{\text{obs},i}^2+\sigma_\text{int}^2},
    \label{eqn:chi2_cosmo_diag_cov}
\end{equation}
where $\mu_\text{cosmo}(z_i;\bm{\Theta})$ is the distance modulus-redshift relation determined by the cosmological parameters $\bm{\Theta}$, and $\sigma_{\text{obs}, i}$ is the observational uncertainty of the measurements. $\sigma_\text{int}$ is the intrinsic dispersion of standardized magnitudes, usually found by iteratively calculating the value of $\sigma_\text{int}$ that ensures the minimum value of $\chi^2$ is equal to 1. This process is effectively a familiar linear regression.

The need to add an additional uncertainty term in the form of $\sigma_\text{int}$ in Equation \ref{eqn:chi2_cosmo_diag_cov} suggests that the linear relationship between SALT2 parameters and absolute magnitude may not capture all of the variation in supernova magnitudes, or that the parametrization provided with SALT2 may not capture all of the information that is needed to fully standardize supernova magnitudes \citep{saunders_snemo_2018}. This motivates the search for other observable properties of SNe~Ia that might explain this remaining variation, as well as the use of these other properties for standardization. One way to search for such properties is to measure correlations between these properties and the Hubble residuals $\mu_i(z_i;m_B^*, x_1, c)-\mu_\text{cosmo}(z_i;\bm{\Theta})$. A number of studies \citep{kelly_hubble_2010, lampeitl_effect_2010, sullivan_dependence_2010, childress_host_2013} have observed such a correlation with the host galaxy stellar mass: supernovae in galaxies with $\log(\mathcal{M}/\mathcal{M}_\odot) > 10$ are $\sim0.1$ magnitudes brighter after standardization than supernovae in galaxies with $\log(\mathcal{M}/\mathcal{M}_\odot) < 10$. \cite{rigault_evidence_2013}, \cite{childress_ages_2014}, and \cite{rigault_confirmation_2015} show that this effect could be due to similar correlations with host galaxy age. However, the significance of some of these correlations has been debated \citep[e.g.][]{jones_reconsidering_2015, jones_should_2018}, indicating that care must be taken in making these measurements.

Reporting the size of correlations with the linear regression residuals is mathematically well-motivated if the covariate used to predict these residuals is not itself correlated with those used in the original regression (if, for example, host mass were not correlated with light curve parameters). However, if this key assumption is violated, we find ourselves in a situation referred to in the statistics and econometrics literature as multicollinearity \cite[e.g.][]{farrar_multicollinearity_1967}. Multicollinearity results in unreliable and biased estimates of effect sizes.  A related concern discussed frequently in these fields is omitted variable bias, in which misspecification of the regression problem results in biased estimates of the true regression parameters \citep{clarke_phantom_2005, wooldridge_introductory_2013}. We can see an example of this in \cite{rigault_strong_2018}, where two different values of the size of the luminosity difference between supernovae in environments with differing star-formation rates are found when measuring the step with a sequential regression versus a simultaneous regression.

This effect can also be seen in studies that use more sophisticated analysis techniques, like the BEAMS with Bias Corrections (BBC) technique of \citet{kessler_correcting_2017}. These methodologies begin by simulating the effects of changing populations of light curve parameters with redshift, selection effects, and contamination by core-collapse supernovae that were incorrectly typed by photometric classification algorithms. The results of these simulations are used to generate additional redshift-dependent corrections to the Hubble diagram and this corrected Hubble diagram is used to constrain cosmological parameters. The regression bias we discuss here would not be present in the results of these studies as long as the covariances between all parameters are included in the analysis. Indeed, \cite{smith_first_2020} found that the underestimate of the host galaxy mass step presented in \cite{brout_first_2019} could be rectified by including the correlation between host galaxy mass and SALT2 $x_1$ in the bias correction framework.

This work aims to explore and quantify the general impact of the non-simultaneous regression methodology used in some Type Ia supernova analyses on reported effect sizes for both linear and step-function residual trends when multicollinearity exists. In Section \ref{sec:toy_model}, we work through an example using a generalized two-dimensional linear regression problem with correlated covariates. In Section \ref{sec:add_step}, we analyze a similar model that includes a step function and compare the results to those obtained in the linear case. We then calculate the effect that this general mathematical model has in the particular case of measuring the host galaxy mass step using literature data of SALT2 parameters and host galaxy masses in Section \ref{sec:data_comparison} and conclude in Section \ref{sec:conclusion} by identifying previous results that have overlooked this effect and recommending that future analyses use fully simultaneous regression techniques.

\section{Toy Model: Two-dimensional Linear Regression with Correlated Covariates} \label{sec:toy_model}
We consider the following toy model: A series of $n$ observations\footnote{Note that $x_1$ here is just the vector of observations in the first dimension and should not be confused with the $x_1$ parameter of the SALT2 model discussed  in Equation \ref{eqn:salt_mu}.} $\{(x_1^{(1)}, x_2^{(1)}), \cdots, (x_1^{(n)}, x_2^{(n)})\}$ is drawn from a two-dimensional Gaussian distribution with $\mu=(0, 0)$ and a covariance matrix given by
\begin{equation}
    \Sigma = \left(
    \begin{matrix}
        \sigma_1^2 & \rho\sigma_1\sigma_2\\
        \rho\sigma_1\sigma_2 & \sigma_2^2
    \end{matrix}
    \right)
    \label{eqn:covmat}
\end{equation}
$\sigma_1$ and $\sigma_2$ are the standard deviations of the observations in the $x_1$ and $x_2$ dimensions, respectively, and $\rho$ is the Pearson correlation coefficient between them. They are not measurement errors, but measures of the natural spread in the distributions.
We then define
\begin{equation}
    y_i=\beta_1 x_1^{(i)} + \beta_2 x_2^{(i)} + \epsilon^{(i)}
\label{eqn:linear_model}
\end{equation}
where $\beta_1$ and $\beta_2$ are the regression coefficients, and $\epsilon$ is a noise vector drawn from a univariate normal distribution $\mathcal{N}(0, \sigma_\text{int}^2+\sigma_\text{obs}^2)$. This noise vector represents a combination of the intrinsic scatter in the model, as well as the observational measurement error. We can reformulate this as a matrix equation by denoting the data matrix as $\bm{X} = (\bm{x}_1, \bm{x}_2)$ and the coefficient vector as $\bm{\beta}=(\beta_1, \beta_2)$, giving $\bm{Y}=\bm{X\beta}+\bm{\epsilon}$.

Standard simultaneous two-dimensional least-squares regression gives us the estimated coefficient vector $\bm{\hat{\beta}}$ which minimizes the square of the residuals between the values predicted by the model ($\bm{\hat{Y}} \equiv \hat{\beta}_1\bm{x}_1 + \hat{\beta}_2\bm{x}_x \equiv \bm{X}\hat{\beta}$) and the data. As we show in Appendix \ref{app:simultaneous_ols}, these estimated values are
\begin{equation}
    \bm{\hat{\beta}} = (\bm{X}^T\bm{X})^{-1}\bm{X}^T(\bm{X\beta} + \bm{\epsilon})
    \label{eqn:sim_beta_vec}
\end{equation}
Since the expectation value of $\bm{\epsilon}$ is 0 by definition, the expectation value of the recovered coefficients from simultaneous regression is identical to the coefficients ($\langle\bm{\hat{\beta}}\rangle=\bm{\beta}$) regardless of the values of the regression coefficients, the covariance matrix components, or the size of the intrinsic scatter. We also show in Appendix \ref{app:simultaneous_ols} that the standard deviation of the residuals ($\bm{r}\equiv\bm{\hat{Y}}-\bm{Y})$ is simply $\sqrt{\sigma_\text{int}^2 + \sigma_\text{obs}^2}$.

In summary, when treating this data set with a simultaneous linear regression, we can reliably recover both the true regression coefficients and intrinsic dispersion. Though there is some uncertainty on the values of the regression coefficients that does depend on the correlation between the covariates, this uncertainty is also inversely proportional to the number of samples fit in the regression and is therefore able to be controlled in the case where $N$ is sufficiently large (see Equation \ref{eqn:var_betahat_simultaneous} in Appendix \ref{app:simultaneous_ols}).

However, oftentimes in Type Ia supernova studies, we do not perform a full simultaneous fit of all of our regression parameters when including additional corrections. Instead, we fit the distance modulus as a linear function of SALT2 parameters and then add a correction to these distance moduli by fitting the distance modulus residuals as a function of some other parameter. This is conceptually analogous to performing this multivariate linear regression one covariate at a time.

We will show that in this case, no biases are introduced if there there is no correlation between the parameters used in the first regression and second regressions (i.e. $\rho=0$). However, if there is some correlation, we find that both the regression coefficients and the estimated scatter on the residuals are biased.

We introduce the notation we will use to treat this situation in our toy example. Without loss of generality, we can first fit $\bm{Y}$ as a function of $\bm{x}_1$. The estimate of the slope will be denoted $\hat{\beta_1}^\prime$ (the prime serves to differentiate this value from the coefficient estimated from the full two-dimensional regression). The residuals of this regression will be denoted $\bm{r}_1$. We then perform a second regression, predicting the residuals of the first regression $\bm{r}_2$ as a function of $\bm{x}_2$. The slope in this case will similarly be denoted $\hat{\beta_2}^\prime$, and the residuals will be denoted by $\bm{r}_2$.

In Appendix \ref{app:non_simultaneous_ols}, we obtain the forms of the expectation values for the regression coefficients resulting from this process, finding that
\begin{equation*}
    \langle\hat{\beta}_1^\prime\rangle = \beta_1 + \frac{\beta_2\rho\sigma_2}{\sigma_1}\quad\text{and}\quad\langle\hat{\beta}_2^\prime\rangle = \beta_2 - \beta_2\rho^2.
\end{equation*}

As we can see, both slopes are biased if $\rho \neq 0$. The size of the bias on both parameters is proportional to the size of the effect and the correlation between the covariates. Additionally, we can recognize that the bias on the first slope is identical to the omitted variable bias. This is expected, as performing this first part of the non-simultaneous regression perfectly simulates the textbook situation presented to describe the omitted variable bias.

We also calculate the spread of the final residuals in Appendix \ref{app:non_simultaneous_ols}, finding
\begin{equation}
    \sigma_{\bm{r}_2}^2 = \beta_2\rho^2\sigma_2^2(1-\rho^2) + \sigma_\text{int}^2 + \sigma_\text{obs}^2
\end{equation}
The standard deviation on the residuals from this analysis, often reported as the root-mean-squared (RMS) residuals, is in fact inflated by a value that scales quadratically with the correlation between the parameters and linearly with the size of the secondary effect. This bias is maximized for a given slope when $\rho = \sqrt{1/2} \approx 0.707$.
\section{Step Function Corrections}
\label{sec:add_step}
Many common analyses used in supernova cosmology do not use a linear model to correct the Hubble diagram residuals for host mass; they use a step function, motivated by the evolution of host galaxy stellar populations with redshift.\footnote{In order to maintain the differentiability of this function, some analyses approximate a step function with a logistic function with a large growth rate.  To ease our calculations (particularly in calculating the expected covariance in Equation \ref{eqn:exp_val_abs_val}), we use the sign function. The differences between the two are negligible for our purposes.} We'll modify the toy model presented in Section \ref{sec:toy_model}, and consider instead
\begin{equation}
    y_i = \alpha x_1^{(i)} + \frac{\gamma}{2}\sgn(x_2^{(i)})
\label{eqn:linear_and_step}
\end{equation}

In the simultaneous case, the expected values of the best-fit regression coefficients $\hat{\alpha}$ and $\hat{\gamma}$ are equivalent to the true values. The proof of this is very similar to the proof for the bilinear toy model presented in Appendix \ref{app:simultaneous_ols}, so we do not present any further details here.

In Appendix \ref{app:step_func}, we have worked through the non-simultaneous case where we fit the linear relationship first, followed by the step function correction to the resulting residuals. The expectation value of the best-fit linear slope ($\hat{\alpha}^\prime$) is
\begin{equation}
    \langle\hat{\alpha}^\prime\rangle = \alpha + \frac{\gamma\rho}{\sigma_1\sqrt{2\pi}}
    \label{eqn:slope_inflation}
\end{equation}
The expected step size obtained from the residuals after correcting for the linear relationship is
\begin{equation}
    \langle\hat{\gamma}^\prime\rangle = \gamma\left(1 - \frac{2\rho^2}{\pi}\right),
    \label{eqn:step_inflation}
\end{equation}
and the spread of the remaining residuals is
\begin{equation}
    \sigma_{\bm{r}_\beta}^2 = \frac{\gamma^2\rho^2}{2\pi}\left(1-\frac{2\rho^2}{\pi}\right) + \sigma_\text{int}^2 + \sigma_\text{obs}^2
\end{equation}
So, using a step-function secondary correction gives us similar biases to the linear secondary correction. The size of the step is underestimated by a factor that scales quadratically with the correlation coefficient between covariates and linearly with the true step size. Additionally, the size of the linear correction term is overestimated by a factor that scales linearly with the step size and the correlation coefficient. Finally, the variance of the residuals after correction is inflated by a factor that scales similarly. The bias on the variance of the residuals is maximal when $\rho=\sqrt{\pi/4}\approx 0.886$.

\section{Comparison to Data}
\label{sec:data_comparison}
The actual distributions of $x_1$, $c$, and $\mathcal{M}_\text{host}$ found in the data are not purely Gaussian, as they are in our toy models. This means that we can no longer derive closed-form relations describing the impact of non-simultaneous fitting. We can, however, simulate the effects. We do so using the published values of $x_1$, $c$, and $\log(\mathcal{M}_\text{host}/\mathcal{M}_\odot)$ from the low- and mid-redshift samples of supernovae from the first three years of the Dark Energy Survey \cite[][hereafter referred to as the Low-$z$ and DES subsamples]{abbott_first_2019}, along with the Pantheon data set \citep{scolnic_complete_2018}, which combines spectroscopically-classified supernovae from PanSTARRS supernovae \cite[PS1;][]{rest_cosmological_2014, scolnic_color_2014} with supernovae from the SuperNova Legacy Survey \cite[SNLS;][]{conley_supernova_2011, sullivan_snls3_2011} and the Sloan Digital Sky Survey \cite[SDSS;][]{frieman_sloan_2008, kessler_first-year_2009, sako_data_2018}.\footnote{The DES and Low-$z$ sample data can be downloaded at \url{https://des.ncsa.illinois.edu/releases/sn}, and the Pantheon data may be found at \url{https://archive.stsci.edu/prepds/ps1cosmo/index.html}.} Each of these data sets shows a fairly strong correlation between $x_1$ and host mass, as seen in Table \ref{tab:corr_coefs}, so we can expect to find non-simultaneous regression biases.

\begin{table}[htbp]
    \centering
    \begin{tabular}{cccc}\toprule
        Data set & $\rho_{x_1, c}$ & $\rho_{x_1, \text{mass}}$ & $\rho_{c, \text{mass}}$\\\midrule
        DES & $-0.087$ & $-0.371$ & $0.1811$\\
        PS1 & $-0.041$ & $-0.248$ & $0.0610$\\
        SDSS & $-0.035$ & $-0.297$ & $0.0002$\\
        SNLS & $0.016$ & $-0.304$ & $0.0629$\\
        Low-$z$ & $0.130$ & $-0.347$ & $-0.1052$\\
        \bottomrule
    \end{tabular}
    \caption{Pearson correlation coefficients between SALT2 parameters and host galaxy masses (measured as $\log(\mathcal{M}_\text{host}/\mathcal{M}_\odot)$). Each data set shows a relatively strong correlation between $x_1$ and mass, indicating that biases can be introduced from non-simultaneous regression.}
    \label{tab:corr_coefs}
\end{table}

To simulate the magnitude of these effects with the non-Gaussian distributions found in the data, we begin by modeling $\delta$, a quantity akin to the Hubble residuals without any corrections for the light curve shape or color parameters and assuming a fixed cosmology:
\begin{equation}
    \delta= \alpha x_1 - \beta c + \frac{\gamma}{2}\sgn\left[\log\left(\frac{\mathcal{M}_\text{host}}{\mathcal{M}_\odot}\right) - 10 \right] + \epsilon
\end{equation}
$\epsilon$ here is a Gaussian distributed noise vector with variance $\sigma_\text{noise}^2$. We use each data set's values of $x_1$, $c$, and $\mathcal{M}_\text{host}$ to calculate 50 instances of $\delta$ for each of approximately 12,000 different combinations of the standardization parameters $\alpha$, $\beta$, $\gamma$, and $\sigma_\text{int}$ in the ranges listed in Table \ref{tab:sim_ranges}. We are motivated to simulate various combinations of the regression coefficients and noise levels by our toy model, which showed that each of these values is intrinsically linked to the others. With each of these data sets, we performed a fully simultaneous regression of all parameters, as well as a non-simultaneous regression of the host galaxy mass correction parameter $\gamma$ after a simultaneous regression of the light curve parameters $\alpha$ and $\beta$, in order to infer the size of the biases.

\begin{table}
    \centering
    \begin{tabular}{cc}
    \toprule
        Parameter & Range \\\midrule
        $\alpha$ & $(0.05, 0.25)$ \\
        $\beta$ & $(2.5, 3.5)$ \\
        $\gamma$ & $(-0.1, 0.1)$ \\
        $\sigma_\text{noise}$ & $(0, 0.2)$ \\
        \bottomrule
    \end{tabular}
    \caption{Ranges for the standardization hyperparameters used in the simulation analysis.}
    \label{tab:sim_ranges}
\end{table}

Each of these simulations gives us a table of true values of $\alpha$, $\beta$, and $\gamma$ (used in the calculation of $\delta$), the simultaneous best-fit values $\hat{\alpha}$, $\hat{\beta}$, and $\hat{\gamma}$, as well as the non-simultaneous best-fit values $\hat{\alpha}^\prime$, $\hat{\beta}^\prime$, and $\hat{\gamma}^\prime$. Regardless of true parameter values, the simultaneous fit parameters all match the true parameters. The magnitude of the error on the non-simultaneous best-fit parameters depends on the data subset in question, because of the differing distributions of the light curve parameters and host galaxy masses, as well as on the true values of the regression parameters $\alpha$, $\beta$, and $\gamma$. We note that the relationships between the true standardization parameters and the non-simultaneously-determined standardization parameters are all linear, i.e.
\begin{equation}
    \gamma = c_{\gamma, 0} + \displaystyle\sum_{i\in\{\hat{\alpha}^\prime, \hat{\beta}^\prime, \hat{\gamma}^\prime\}} c_{\gamma, i}i,
    \label{eqn:lin_decomp_reg_bias}
\end{equation}
where the $c$ values are linear coefficients that can be measured from the simulation data sets.\footnote{These coefficients are not to be confused with the SALT2 color parameter $c$ (with no subscript).} Similar relationships exist for $\alpha$ and $\beta$ as well. This is not unexpected; we can find this linear relationship by rearranging the toy model results of Equations \ref{eqn:slope_inflation} and \ref{eqn:step_inflation}:
\begin{align}
    \alpha &=\hat{\alpha}^\prime - \sqrt{\frac{\pi}{2\sigma_1^2}}\frac{\rho}{1-\rho^2}\hat{\gamma}^\prime &
    \gamma &= \frac{\pi}{1-2\rho^2}\hat{\gamma}^\prime
    \label{eqn:rearranged}
\end{align}
Non-zero values of coefficients other than $c_{x, 0}$ indicate that there is ``leakage" from one standardization parameter to the other; for example, if $c_{\gamma, \hat{\alpha}^\prime} \neq 0$, then the size of the $\alpha$ correction impacts the reported size of the $\gamma$ corrections. Moreover, these coefficients define a linear transformation between the true regression parameters and those coming from a non-simultaneous fit, so the inverse of these transformations can be used to correct previous non-simultaneous regressions. The transformations we obtained from our simulations are presented in Table \ref{tab:trans}, where we have omitted columns that are expected to be 0 (e.g. $c_{\alpha,\hat{\beta}^\prime}$) or 1 (e.g. $c_{\alpha, \hat{\alpha}^\prime}$).

\begin{table}[htbp]
    \centering
    \begin{tabular}{cccc}\toprule
    Data set & $c_{\alpha,\hat{\gamma}^\prime}$ & $c_{\beta,\hat{\gamma}^\prime}$ & $c_{\gamma,\hat{\gamma}^\prime}$ \\\midrule
    DES     & $0.335$ & $-0.702$ & $1.302$\\
    PS1     & $0.135$ & $-0.607$ & $1.111$\\
    SDSS    & $0.125$ & $-0.134$ & $1.237$\\
    SNLS    & $0.203$ & $-0.565$ & $1.14$\\
    Low-$z$ & $0.194$ & $+1.258$ & $2.072$\\\bottomrule
    \end{tabular}
    \caption{Linear transformation coefficients (see Equation \ref{eqn:lin_decomp_reg_bias}) between the standardization hyperparameter $\alpha$, representing the light curve shape-luminosity correction, obtained with a non-simultaneous fit and the true values.}
    \label{tab:trans}
\end{table}

These values that are found from our simulations are roughly consistent with the closed-form solutions (Equation \ref{eqn:rearranged}). Data sets with the largest correlations between light curve parameters and host galaxy mass also exhibit the most leakage between the corresponding standardization parameters and host mass step size. The remaining differences between the values predicted by Equation \ref{eqn:rearranged} and the values in Table \ref{tab:trans} are due to the non-Gaussianity of the $x_1$, $c$, and $\mathcal{M}_\text{host}$ distributions.

We can see that there is significant leakage between the size of the host mass step and the stretch and color standardization parameters $\alpha$ and $\beta$. Multiplying the coefficients relating the non-simultaneously obtained step-size by the typical size of the measured step (0.07 mag.), we can see that this leakage results in a 5--10\% error on the typical size (0.14) of the stretch parameter $\alpha$ and a $\sim$1\% error on the typical size (3.0) of the color parameter $\beta$.

More importantly, the coefficients relating the non-simultaneous step size to the true step size are greater than one for each data set. This means that by fitting the step function separately from other corrections, the true size of the step is underestimated by 10--30\%, and by a factor of two for the Low-$z$ subsample.

The linear transformations presented here can be used directly to correct step sizes obtained from simple non-simultaneous regressions. These simulations do not, however, account for selection effects, redshift-dependent drifts in the populations of the light curve parameters, or models of intrinsic scatter. A number of simulation studies \citep[e.g.][]{scolnic_systematic_2014, scolnic_color_2014, marriner_more_2011, scolnic_measuring_2016} have shown that measurements of the standardization parameters (i.e. $\alpha$ and $\beta$) can be biased by inaccurate specifications of the distributions of the supernova parameters ($x_1$ and $c$). These biases can be corrected using simulations that include these selection effects or avoided using a full Bayesian methodology (as is done in \citet{rubin_unity_2015}, for example). A complete treatment of all of these effects is outside the scope of this work, but the results we've shown here, along with the discussion of the relationship between the host galaxy mass step and bias corrections in \citet{smith_first_2020}, emphasize that a complete supernova cosmology analysis must specify all covariances completely and determine the luminosity correction parameters simultaneously to avoid bias.

\section{Conclusions}
\label{sec:conclusion}
We have worked through a pedagogical example to show that performing linear regression one covariate at a time produces biased estimates of both the regression coefficients and spread of residuals when the covariates are correlated. The sizes of these biases depend directly on the magnitude of the correlation, and there are linear relationships between the error on the estimated slopes and the size of the factor that inflates the estimate of the spread of the remaining scatter. We have proven that similar relationships also hold when fitting step functions to the residuals of a linear regression (as is sometimes done in supernova cosmology) if there are correlations between the parameters being fit in each step. 

We have also presented numerical simulations based on observed data to find corrections to the biases that are introduced from non-simultaneous regression methods. Each data set studied shows the possibility of a large underestimate of the size of the host mass step regardless of values of other nuisance parameters. There are also minor biases in the model parameters governing the relationship between luminosity and light curve width (SALT2 $\alpha$) and luminosity and color (SALT2 $\beta$).

Biases are introduced when the assumptions underlying an analysis method are overlooked. In this particular case, there is an implicit assumption that all covariates must be uncorrelated in order to prevent biases from performing a two-step regression. A number of studies \citep[e.g.][]{kelly_hubble_2010, sullivan_dependence_2010, childress_host_2013, jones_reconsidering_2015, jones_should_2018, rose_think_2019, kelsey_effect_2020} have neglected this effect, leading to underestimated sizes and significances of the effect sizes they report. Both \citet{rigault_strong_2018} and \citet{jones_should_2018} find that simultaneous fits of the light curve and host galaxy standardization parameters result in a larger step size and reduced Hubble residual dispersion, though the latter analysis reports the non-simultaneous fits as its main result. This effect can be entirely explained by the effect we have discussed in this work. 

For the most part, recent cosmology analyses, \citep[e.g.][]{betoule_improved_2014, scolnic_complete_2018}, do properly account for this effect by fitting for the host mass step size simultaneously with the other standardization parameters. It is additionally worth noting that correcting for this bias in reported step sizes is unlikely to resolve the current tension in measurements of the Hubble constant -- a 20\% underestimate of the size of the mass step propagates to $<1$\% change in the value of $H_0$. However, it is not yet clear if the host mass correlations are properly accounted for in the bias corrections, as suggested in \citet{smith_first_2020}. Overall, care must be taken in presenting the size and significance of these relationships and propagating these correlations throughout the analysis. The biases presented here can be easily avoided by fitting all nuisance parameters simultaneously and fully specifying all covariances when presenting measurements of luminosity corrections beyond light curve parameters.


\acknowledgements
The author would like to thank Saul Perlmutter, Greg Aldering, and Ben Rose for comments on this manuscript and helpful discussions. We also thank the anonymous referee for their time and attention. S.D. acknowledges support, in part, from NASA through grant NNG16PJ311I.

\software{
Matplotlib \citep{hunter_matplotlib_2007}, 
Numpy \citep{oliphant_numpy_2006}, 
Pandas \citep{mckinney_data_2010}, 
Python, 
scikit-learn \citep{pedregosa_scikit-learn_2011}, 
SciPy \citep{virtanen_scipy_2020}}


\bibliography{references}
\bibliographystyle{aasjournal}

\appendix
\section{Derivation of Regression Parameters in the Simultaneous Case}
\label{app:simultaneous_ols}
In ordinary least-squares regression, our goal is to minimize the loss function, $L$, defined by the square of the residuals between the $y$ values predicted by our model ($\hat{\bm{Y}}\equiv\hat{\beta}_1\bm{x}_1 +\hat{\beta}_2\bm{x}_2\equiv\bm{X\hat{\beta}}$) and the data. This is equivalent to maximizing the likelihood assuming the residuals are Gaussian.
\begin{align*}
    L &= ||\hat{\bm{Y}}-\bm{Y}||^2\\
    &= (\bm{X\hat{\beta}}-\bm{Y})^T(\bm{X\hat{\beta}}-\bm{Y})\\
    &= \bm{\hat{\beta}}^T\bm{X}^T\bm{X\hat{\beta}}
    - \bm{\hat{\beta}}^T\bm{X}^T\bm{Y}
    - \bm{Y}^T\bm{X\hat{\beta}}
    + \bm{Y}^T\bm{Y}
\end{align*}
We can minimize this by taking the gradient as a function of $\bm{\hat{\beta}}$ and setting it equal to zero.
\begin{align*}
    \frac{\partial L}{\partial\bm{\hat{\beta}}} &=
    (\bm{X}^T\bm{X}\bm{\hat{\beta}})^T
    + \bm{\hat{\beta}}^T\bm{X}^T\bm{X}
    - (\bm{X}^T\bm{Y})^T
    - \bm{Y}^T\bm{X}\\
    &= 2\bm{\hat{\beta}}^T\bm{X}^T\bm{X} - 2\bm{Y}^T\bm{X}
\end{align*}
$$\frac{\partial L}{\partial\bm{\hat{\beta}}} = 0 \Rightarrow \bm{\hat{\beta}}=(\bm{X}^T\bm{X})^{-1}\bm{X}^T\bm{Y}$$
Plugging in our definition of $\bm{Y}$, we get
\begin{equation}
    \bm{\hat{\beta}} = (\bm{X}^T\bm{X})^{-1}\bm{X}^T(\bm{X\beta} + \bm{\epsilon})
\label{eqn:sim_beta_vec_app}
\end{equation}
Since, by definition, $\langle\bm{\epsilon}\rangle=0$, $\langle\bm{\hat{\beta}}\rangle = \beta$, we can show that the spread of the residuals ($\bm{r}\equiv\bm{\hat{Y}}-\bm{Y}$) is simply $\sqrt{\sigma_\text{int}^2 + \sigma_\text{obs}^2}$:
\begin{align*}
    \text{var}(\bm{r}) &= \langle\bm{r}^2\rangle - \langle\bm{r}\rangle^2\\
    &= \langle(\bm{X\hat{\beta}}-\bm{X\beta}-\bm{\epsilon})(\bm{X\hat{\beta}}-\bm{X\beta}-\bm{\epsilon})^T\rangle - \langle(\bm{X\hat{\beta}}-\bm{X\beta}-\bm{\epsilon})\rangle^2\\
    &= \langle\bm{\epsilon}\bm{\epsilon}^T\rangle - \langle\bm{\epsilon}\rangle^2\\
    &= \text{var}(\bm{\epsilon}) = \sigma_\text{int}^2 + \sigma_\text{obs}^2
\end{align*}

The variance on these regression coefficients can also be calculated. First, we calculate $\langle\bm{\hat{\beta}}^2\rangle$:
\begin{align*}
    \langle\bm{\hat{\beta}}^2\rangle &= \langle\bm{\hat{\beta}}\bm{\hat{\beta}}^T\rangle \\
    &= \langle(\bm{X}^T\bm{X})^{-1}\bm{X}^T\bm{YY}^T\bm{X}(\bm{X}^T\bm{X})^{-1}\rangle \\
    &= \langle(\bm{X}^T\bm{X})^{-1}\bm{X}^T(\bm{X\beta}+\bm{\epsilon})(\bm{\beta}^T\bm{X}^T+\bm{\epsilon})\bm{X}(\bm{X}^T\bm{X})^{-1}\rangle\\
    &= \bm{\beta\beta}^T + \sigma_\text{int}^2(\bm{X}^T\bm{X})^{-1}
\end{align*}
Then, by using the definition $\langle\bm{\hat{\beta}}\rangle^2 = \bm{\beta\beta}^T$, we have
$$\text{var}(\bm{\hat{\beta}})= \langle\bm{\hat{\beta}}^2\rangle-\langle\bm{\hat{\beta}}\rangle^2 = (\sigma_\text{int}^2 + \sigma_\text{obs}^2)(\bm{X}^T\bm{X})^{-1}.$$
Calculating the individual components of this variance matrix in our two-dimensional case gives
\begin{equation}
    \text{var}(\hat{\beta_1})=\frac{\sigma_\text{int}^2+\sigma_\text{obs}^2}{N\sigma_1^2\left(1-\rho^2\right)}\quad\text{and}\quad\text{var}(\hat{\beta_2})=\frac{\sigma_\text{int}^2+\sigma_\text{obs}^2}{N\sigma_2^2\left(1-\rho^2\right)}.
    \label{eqn:var_betahat_simultaneous}
\end{equation}

\section{Derivation of Biases on Regression Parameters in the Non-Simultaneous Case}
\label{app:non_simultaneous_ols}
We can modify Equation \ref{eqn:sim_beta_vec} to obtain the predicted value of the slope in the first fit:
\begin{align*}
    \langle\hat{\beta}_1^\prime\rangle &= \langle(\bm{x}_1^T\bm{x}_1)^{-1}\bm{x}_1^T\bm{Y}\rangle\\
    &= \langle(\bm{x}_1^T\bm{x}_1)^{-1}\bm{x}_1^T\bm{x}_1\beta_1 + (\bm{x}_1^T\bm{x}_1)^{-1}\bm{x}_1^T\bm{x}_2\beta_2 + (\bm{x}_1^T\bm{x}_1)^{-1}\bm{x}_1^T\bm{\epsilon}\rangle\\
    &= \beta_1 + \beta_2\langle(\bm{x}_1^T\bm{x}_1)^{-1}\bm{x}_1^T\bm{x}_2\rangle\\
    &= \beta_1 + \frac{\beta_2\rho\sigma_2}{\sigma_1}
\end{align*}

The residuals from this first regression are
\begin{align*}
    \bm{r}_1 &= \bm{Y}-\bm{\hat{Y}}_1 \\
    &= \beta_1\bm{x}_1 + \beta_2\bm{x}_2 + \bm{\epsilon}-\hat{\beta}_1^\prime\bm{x}_1\\
    &= \beta_2\bm{x}_2 - \frac{\beta_2\rho\sigma_2}{\sigma_1}\bm{x}_1 + \bm{\epsilon}
\end{align*}
We can go through a similar analysis to find the predicted secondary effect from fitting the residuals of the first regression $\bm{r}_1$ as a function of $\bm{x}_2$. This gives
\begin{align*}
    \langle\hat{\beta}_2^\prime\rangle &= \langle(\bm{x}_2^T\bm{x}_2)^{-1}\bm{x}_2^T\bm{r}_1\rangle\\
    &= \langle(\bm{x}_2^T\bm{x}_2)^{-1}\bm{x}_2^T(\beta_2\bm{x}_2 - \frac{\beta_2\rho\sigma_2}{\sigma_1}\bm{x}_1 + \bm{\epsilon})\rangle\\
    &= \beta_2 - \beta_2\rho^2
\end{align*}

Calculating the final residuals gives
\begin{align*}
    \bm{r}_2 &= \bm{r}_1 - \bm{\hat{r}}_1\\
    &= \beta_2\bm{x}_2 - \frac{\beta_2\rho\sigma_2}{\sigma_1}\bm{x}_1 + \bm{\epsilon} - \hat{\beta}_2^\prime\bm{x}_2\\
    &= - \frac{\beta_2\rho\sigma_2}{\sigma_1}\bm{x}_1 + \beta_2\rho^2\bm{x}_2 + \bm{\epsilon},
\end{align*}
and using the typical propagation of uncertainty formulae to find the variance of these residuals, we obtain
\begin{align*}
    \sigma_{\bm{r}_2}^2 &= \frac{\beta_2^2\rho^2\sigma_2^2}{\sigma_1^2}\sigma_1^2 + \beta_2^2\rho^4\sigma_2^2 - 2\frac{\beta_2^2\rho^3\sigma_2}{\sigma_1}\rho\sigma_1\sigma_2 + \sigma_\text{int}^2\\
    &= \beta_2^2\rho^2\sigma_2^2\left(1-\rho^2\right) + \sigma_\text{int}^2.
\end{align*}

\section{Step Function Correction Derivations}
\label{app:step_func}
Again, we start with Equation \ref{eqn:sim_beta_vec} to obtain the expected value of the best-fit slope from the linear portion of the fit.
\begin{align*}
    \langle\hat{\alpha}^\prime\rangle &= \langle \bm{x}_1^T\bm{x}_1)^{-1}\bm{x}_1^T\bm{Y}\rangle \\
    &= \langle(\bm{x}_1^T\bm{x}_1)^{-1}\bm{x}_1^T\bm{x}_1\alpha + (\bm{x}_1^T\bm{x}_1)^{-1}\bm{x}_1^T\sgn(\bm{x}_2)\frac{\gamma}{2}+(\bm{x}_1^T\bm{x}_1)^{-1}\bm{x}_1^T\bm{\epsilon}\rangle\\
    &= \alpha + \frac{\gamma}{2\sigma_1^2}\langle\bm{x}_1^T\sgn(\bm{x}_2)\rangle \\
    &= \alpha + \frac{\gamma\rho}{\sigma_1\sqrt{2\pi}}
\end{align*}
The proof of the final step is as follows, where $p(x_1, x_2)$ is a bivariate Gaussian distribution with mean $(0, 0)$ and covariance matrix like that in Equation \ref{eqn:covmat}.
\begin{align}
\label{eqn:exp_val_abs_val}
    \langle \bm{x}_1\sgn(\bm{x}_2)\rangle &= \displaystyle\int_{-\infty}^\infty \displaystyle\int_{-\infty}^\infty x_1\sgn(x_2)p(x_1, x_2)dx_1dx_2\nonumber\\
    &= \displaystyle\int_{-\infty}^0 \displaystyle\int_{-\infty}^\infty -x_1 p(x_1, x_2)dx_1dx_2+
    \displaystyle\int_{0}^\infty \displaystyle\int_{-\infty}^\infty x_1 p(x_1, x_2)dx_1dx_2\nonumber\\
    &= 2 \displaystyle\int_{0}^\infty \displaystyle\int_{-\infty}^\infty x_1 p(x_1, x_2)dx_1dx_2\nonumber\\
    &= \frac{1}{\pi\sigma_1\sigma_2\sqrt{1-\rho^2}}\displaystyle\int_{0}^\infty \displaystyle\int_{-\infty}^\infty x_1 \exp\left[-\frac{1}{2(1-\rho^2)}\left(\frac{x_1^2}{\sigma_1^2}+\frac{x_2^2}{\sigma_2^2}-\frac{2\rho x_1 x_2}{\sigma_1\sigma_2}\right)\right]dx_1dx_2\nonumber\\
    &= \sqrt\frac{2}{\pi}\rho\sigma_1
\end{align}

The residuals that remain after correcting for the linear slope are
\begin{align*}
    \bm{r}_\alpha &= \bm{Y} - \bm{\hat{Y}}_\alpha\\
    &= \alpha\bm{x}_1 + \frac{\gamma}{2}\sgn(\bm{x}_2) +\bm{\epsilon} - \hat{\alpha}^\prime\bm{x}_1\\
    &= \frac{\gamma}{2}\sgn(\bm{x}_2) - \frac{\gamma\rho}{\sigma_1\sqrt{2\pi}}\bm{x}_1 + \bm{\epsilon}
\end{align*}
We can find what the step size $\gamma$ would be when fit to these residuals by finding the value of $\hat{\gamma}^\prime$ that minimizes $L=\left\|\bm{r}_\alpha - \frac{\hat{\gamma}^\prime}{2}\sgn(\bm{x}_2)\right\|^2$.
\begin{align*}
    L &= \left\|\bm{r}_\alpha - \frac{\hat{\gamma}^\prime}{2}\sgn(\bm{x}_2)\right\|^2\\
    &= \bm{r}_\alpha^2 - \hat{\gamma}^\prime\bm{r}_\alpha\sgn(\bm{x}_2) + \frac{\hat{\gamma}^{\prime 2}}{4}\\
    \frac{\partial L}{\partial \hat{\gamma}^\prime} &= -\bm{r}_\alpha\sgn(\bm{x}_2) + \frac{\hat{\gamma}^\prime}{2}
\end{align*}
Setting this derivative to zero, we find
\begin{align*}
    \hat{\gamma}^\prime &= 2\bm{r}_\alpha\sgn(\bm{x}_2)\\
    &= \gamma - \frac{2\gamma\rho}{\sigma_1\sqrt{2\pi}}\bm{x}_1\sgn(\bm{x}_2) + 2\bm{\epsilon}\sgn(\bm{x}_2)
\end{align*}
The expectation value is
\begin{align*}
    \langle\hat{\gamma}^\prime\rangle &= \gamma - \frac{2\gamma\rho}{\sigma_1\sqrt{2\pi}}\langle\bm{x}_1\sgn(\bm{x}_2)\rangle + 2\langle\bm{\epsilon}\sgn(\bm{x}_2)\rangle\\
    &= \gamma - \frac{2\gamma\rho^2}{\pi}
\end{align*}
where we used the result of Equation \ref{eqn:exp_val_abs_val} to evaluate $\langle\bm{x}_1\sgn(\bm{x}_2)\rangle$. Our final residuals after the two-step regression are then
\begin{align*}
    \bm{r}_\beta &= \bm{r}_\alpha - \bm{\hat{r}}_\alpha \\
    &= \frac{\gamma}{2}\sgn(\bm{x}_2) - \frac{\gamma\rho}{\sigma_1\sqrt{2\pi}}\bm{x}_1 + \bm{\epsilon} - \frac{\gamma}{2}\sgn(\bm{x}_2) + \frac{\gamma\rho^2}{\pi}\sgn(\bm{x}_2)\\
    &= -\frac{\gamma\rho}{\sigma_1\sqrt{2\pi}}\bm{x}_1+\frac{\gamma\rho^2}{\pi}\sgn(\bm{x}_2)+\bm{\epsilon}
\end{align*}
The variance of these residuals is
\begin{align*}
    \sigma_{\bm{r}_\beta}^2 &= \frac{\gamma^2\rho^2}{2\pi\sigma_1^2}\sigma_1^2 + \frac{\gamma^2\rho^4}{\pi^2} - \frac{2\gamma^2\rho^3}{\sigma_1\sqrt{2\pi^3}}\langle\bm{x}_1\sgn(\bm{x}_2)\rangle + \sigma_\text{int}^2\\
    &= \frac{\gamma^2\rho^2}{2\pi} + \frac{\gamma^2\rho^4}{\pi^2} - \frac{2\gamma^2\rho^4}{\pi^2} + \sigma_\text{int}^2\\
    &= \frac{\gamma^2\rho^2}{2\pi}\left(1-\frac{2\rho^2}{\pi}\right) + \sigma_\text{int}^2
\end{align*}


\end{document}